\documentclass[12pt]{article}
\usepackage{graphicx}
\usepackage{natbib}
\usepackage{url} 

\newcommand{\blind}{0}

\usepackage{amsfonts, amsmath, mathrsfs, bm}
\usepackage{float, multirow, booktabs, threeparttable}
\usepackage{array}
\usepackage{tikz}
\usepackage{verbatim}

\usepackage{color}

\newtheorem{theorem}{Theorem}
\newtheorem{proposition}{Proposition}
\newtheorem{remark}{Remark}

\newcommand{\PreserveBackslash}[1]{\let\temp=\\#1\let\\=\temp}
\newcolumntype{C}[1]{>{\PreserveBackslash\centering}p{#1}}
\newcolumntype{R}[1]{>{\PreserveBackslash\raggedleft}p{#1}}
\newcolumntype{L}[1]{>{\PreserveBackslash\raggedright}p{#1}}

\newcommand{\convergeto}{\overset{d}{\longrightarrow}}
\def\T{{ \mathrm{\scriptscriptstyle \top} }}
\newcommand{\bas}{\begin{eqnarray*}}
\newcommand{\eas}{\end{eqnarray*}}

\newcommand{\ba}{\begin{eqnarray}}
\newcommand{\ea}{\end{eqnarray}}

\newcommand{\bit}{\begin{itemize}}
\newcommand{\eit}{\end{itemize}}

\newcommand{\ben}{\begin{enumerate}}
\newcommand{\een}{\end{enumerate}}

\newcommand{\e}{ { \mathbb{E}}}

\newcommand{\pr}{\operatorname{pr}}

\newcommand{\bbeta}{ {\bm \beta} }

\newcommand{\btheta}{ {\bm \theta}}
\newcommand{\bpsi}{ {\bm \psi}}
\newcommand{\bzero}{ {\bf 0}}

\newcommand{\bX}{ {\bf X}}

\newcommand{\bx}{ {\bf x}}
\newcommand{\bz}{ {\bf z}}

\newcommand{\be}{ {\bf e}}
\newcommand{\bW}{ {\bf W}}

\newcommand{\bV}{ {\bf V}}

\newcommand{\mI}{ { \mathcal{I} }}
\newcommand{\mQ}{ { \mathcal{Q} }}

\newcommand{\rM}{ {\rm M}}

\newcommand{\rd}{ {\rm d}}
\newcommand{\rBi}{ {\rm Bi}}
\newcommand{\rN}{ {\rm N}}

\usepackage{hyperref}

\begin{document}

\def\spacingset#1{\renewcommand{\baselinestretch}%
{#1}\small\normalsize} \spacingset{1}


\if0\blind
{
  \title{\bf Penalized empirical likelihood estimation and EM algorithms
for closed-population capture--recapture models}
  \author{Yang Liu\hspace{.2cm}\\
    KLATASDS -- MOE, School of Statistics, East China Normal University, \\
    Shanghai 200062, China
    \\
    and \\
    Pengfei Li \\
    Department of Statistics and Actuarial Science, University of Waterloo, \\
    Ontario N2L 3G1, Canada
    \\
    and \\
    Yukun Liu\thanks{Corresponding author: Yukun Liu (Email: ykliu@sfs.ecnu.edu.cn).} \\
    KLATASDS -- MOE, School of Statistics, East China Normal University, \\
    Shanghai 200062, China
    }
    \date{}
  \maketitle
} \fi


\if1\blind
{
  \bigskip
  \bigskip
  \bigskip
  \begin{center}
    {\LARGE\bf Penalized empirical likelihood estimation and EM algorithms
for closed-population capture--recapture models}
\end{center}
  \medskip
} \fi

\vspace{-1em}

\bigskip
\begin{abstract}
Capture--recapture experiments are widely used to estimate
the abundance of a finite population.
Based on capture--recapture data, the empirical likelihood (EL) method
has been shown to outperform the conventional
conditional likelihood (CL) method.
However, the current literature on EL abundance estimation
ignores behavioral effects,
and the EL estimates may not be stable,
especially when the capture probability is low.
We make three contributions in this paper.
First, we extend the EL method to capture--recapture models that account for behavioral effects.
Second, to overcome the instability of the EL method,
we propose a penalized EL (PEL) estimation method
that penalizes large abundance values.
We then investigate the asymptotics of the maximum PEL estimator and the PEL ratio statistic.
Third, we develop 
standard expectation--maximization (EM) algorithms
for PEL to improve its practical performance.
The EM algorithm is also applicable to EL and CL with slight modifications.
Our simulation and a real-world data analysis demonstrate that the PEL method successfully overcomes the instability of the EL method
and the proposed EM algorithm produces more reliable
results than existing optimization algorithms.
\end{abstract}

\noindent%
\vfill

\spacingset{1.5} 

\section{Introduction}
\label{s:intro}

The abundance or size of a finite population is of great importance
in many fields, such as fishery, ecology, demographics, and epidemiology
\citep{Bohning2018capture}.
For example, fish abundances are fundamental for
evaluating fishery resources
and bird abundances are important indices
for planning and assessing habitat conservation policies.
This paper considers closed populations,
in which there are no births, deaths, or migration,
so that the population size is a constant over
the time period of the study.
The closed-population assumption is reasonable in most cases
since the study period is usually short.
Capture--recapture experiments are widely used to estimate the abundance of a closed population, as this is a cost-effective technique for collecting data.
In such experiments, individuals from the population of interest are
captured, marked, and then released.
At a later time, after the captured individuals have mixed with
other members of the population, another sample is taken.

Capture--recapture experiments can be discrete-time or continuous-time,
according to whether the capture efforts are made for a finite number of discrete occasions
or over a particular period of time.
Regardless of whether a capture--recapture experiment is discrete-time or continuous-time,
the probability or intensity of capture is often influenced by
three factors: individual heterogeneity, time, and the behavioral response to capture
 \citep{otis1978statistical}.
It is natural that different individuals have different capture probabilities
since they have different covariates (individual heterogeneity),
such as sex for the black bear population analyzed in Section~\ref{sec:data}.
The time factor implies that an individual may have different capture probability on each capture occasion.
In addition, individuals may make different behavioral responses to
different types of capture equipment, leading to different capture probabilities.
For example, black bears might develop a bait preference after being captured,
and become so ``happy'' that they are more easily captured again.
In contrast, cliff swallows are often caught by mist nets \citep{roche2013recapture}, so might tend subsequently to avoid nets and feel so ``sad'' that they are less easily recaptured.
Throughout this paper, we restrict our
attention to the discrete-time capture--recapture data
and similar conclusions can be straightforwardly extended to the
continuous-time capture--recapture data.

Any combination of the aforementioned three factors can be used to formulate
a specific type of probability 
model \citep{otis1978statistical};
see Tables~\ref{tab:model}.
Of all the possible models,
the Huggins--Alho model 
is the most popular for discrete-time
capture--recapture experiments. 
In these models, conditional likelihood (CL) \citep{huggins1991some} and
empirical likelihood (EL) \citep{liu2017maximum}
are the two most common ways to estimate abundance.
The CL method has two steps.
A maximum CL estimator is first obtained
for the underlying nuisance model parameters,
and then a Horvitz--Thompson type estimator
and Wald-type confidence intervals
are constructed for the abundance.
However, if the estimated capture probabilities are small,
the CL-based Horvitz--Thompson estimator for abundance
is unstable and
the corresponding Wald-type confidence intervals may
have unconvincing widths or even infinite upper limits.
The EL abundance estimation method, proposed by \cite{liu2017maximum}
for discrete-time capture--recapture experiments,
yields
a more stable point estimator
and a more accurate interval estimator than the CL method.
It has been extended to continuous-time capture--recapture experiments
\citep{liu2018full} and to cases with missing covariates \citep{liu2021maximum}.
Nevertheless, if the marginal capture probability is
quite small, the EL method also becomes unstable and may
produce poor estimates.
These features are illustrated in our simulation results in Section~\ref{sec:sim}
and the black bear data analysis in Section~\ref{sec:data}.

The EL literature on capture--recapture data
takes the effects of time and individual heterogeneity into account
but largely ignores the effects of different behavioral responses.
Behavioral heterogeneity has attracted much attention in
the CL literature.
\cite{huggins1991some} and
\cite{chao2000capture} considered that behavioral response could be enduring
and proposed a CL estimation method.
An enduring behavioral response means that
the capture probabilities are different but remain constant before and after
one individual is captured.
Special cases include the ``sad'' and ``happy'' behavioral effects.
As noted by \cite{xi2009estimation}, the estimates may be quite unreliable
if some major variables affecting the capture
probability are not included in the model.
Hence, it is necessary to develop EL methods
for capture--recapture models that account for behavioral heterogeneity.

Besides the estimation methods themselves,
there is an urgent need to improve the algorithms that
implement the CL and EL methods.
The {\tt R} package {\tt VGAM} developed by \cite{yee2015vgam}
seems an appealing tool for implementing the CL method
for discrete-time capture--recapture studies.
Compared with the generalized linear model (GLM) classes,
the vector GLM classes used there limit the applicability of {\tt VGAM}.
As the profile log EL function depends on
a Lagrange multiplier, which is an implicit function,
the EL method involves double optimizations.
Its implementation is usually achieved via {\tt R} built-in optimization functions
\citep{liu2017maximum,liu2018full,liu2021maximum},
although the resulting solutions depend on the choice of initial values.
The expectation--maximization (EM) algorithm \citep{dempster1977maximum}
is well known for its stable performance
and several EM-like algorithms have been adapted to capture--recapture models
\citep{wang2005semiparametric,xi2009estimation,farcomeni2013heterogeneity,farcomeni2016general}.
Unfortunately, there is no guarantee
for these algorithms that the iterations will converge to maximum likelihood estimators
or that the likelihood will increase monotonically after each step,
which are the most appealing properties of the standard EM algorithm.

The aforementioned imperfections
of the existing EL approaches for capture--recapture data motivate our work,
which has three main contributions.
First, we extend the EL method of \cite{liu2017maximum}
to general capture--recapture models, especially those
accounting for behavioral heterogeneity.

Second, to overcome the instability of the EL method,
we propose a penalized EL (PEL) approach based on the EL method
that penalizes large abundance values.
With an appropriately chosen data-adaptive penalty function,
the maximum PEL estimator has a normal distribution
and the PEL ratio statistic has a central chi-squared distribution asymptotically.
The penalty term shrinks the maximum EL estimator toward
\cite{chao1987estimating, chao1989estimating}'s lower bound,
making the maximum PEL estimator much more stable than the maximum EL estimator.
Also, it quickly pushes the PEL ratio function to infinity for large abundance values. This allows the PEL method to overcome the possible flatness issue of the right tail of the EL ratio function, so that it produces better interval estimators with convincing upper limits, especially when the marginal capture probability is small.
These desirable properties of the PEL method are confirmed by our simulation results.

Third, we develop a series of standard EM algorithms
for the PEL method for various capture--recapture models.
These algorithms retain the nice increasing property of
the PEL after each iteration.
They are also applicable to the CL and EL methods
with slight modifications.
An appealing property of these EM algorithms
is that one key optimization
can easily be implemented with standard generalized linear regression programs.
Our numerical studies show that
the EM algorithms proposed are more flexible and
produce more reliable results than
the existing optimization algorithms.

The rest of this article is organized as follows.
In Section~\ref{sec:like},
we extend the EL method to general capture--recapture models,
introduce the proposed PEL method,
and establish its asymptotic properties.
In Section~\ref{sec:alg}, we develop the standard EM algorithms
and investigate their finite-sample properties.
Our simulation results and an analysis of a real-world data set
are provided in Sections~\ref{sec:sim} and~\ref{sec:data}, respectively.
Section~\ref{sec:con} concludes.
For clarity, the technical proofs are in the online supplementary material.

\section{PEL inference}
\label{sec:like}


Let $N$ be the size of the population of interest
and $K$ the number of capture attempts made
to collect data.
For a generic individual in the population,
we let $\bX$ denote its covariate
with cumulative distribution function $F(\bx)$.
Let $(D_{(1)}, \dots, D_{(K)} )^\T \in \{0, 1\}^{K}$ be
its capture history,
where $D_{(k)}=1$ if the individual is captured
on the $k$th occasion and 0 otherwise.
Given $\bX = \bx$ and the first $(k-1)$ capture statuses,
the conditional probability of the individual being captured
on the $k$th capture occasion is often characterized
by a linear logistic model:
\begin{equation}
\begin{aligned}
\label{eq:cap-prob}
P(D_{(k)} = 1 \mid \bX = \bx,
D_{(1)} = d_{(1)},\dots, D_{(k - 1)} = d_{(k-1)})
  = \frac{\exp(\bbeta^\T \bz_k)}{1 + \exp(\bbeta^\T \bz_k)} =:
g(\bz_k; \bbeta)
\end{aligned}
\end{equation}
for $k=1,\dots,K$,
where $\bz_k$ is
a summarized quantity of the vector
$(\bx^\T, d_{(1)}, \dots, d_{(k-1)})^\T$
and $\bbeta$ is an unknown vector-valued parameter.
Here, we assume that $d_{(0)}=0$, as no capture occurs
before the first capture occasion.
This model is the well-known Huggins--Alho model
\citep{huggins1989statistical, alho1990logistic}.
With different choices of $\bz_k$ and $\bbeta$,
model~\eqref{eq:cap-prob} covers all the eight possible types of
capture--recapture models \citep{otis1978statistical}
that account for time ($t$), individual heterogeneity ($h$),
and/or behavioral response ($b$). See Table~\ref{tab:model}.
The last column $\bz_{k0}$ has the same form as $\bz_{k}$ except that the $d_{(k)}$'s are zeros.

\begin{table}[bt!]
\centering
\caption{Capture probability models.\label{tab:model}}
\begin{threeparttable}
	\begin{tabular}{L{2cm}L{4cm}L{4cm}L{3cm} @{}}
	\toprule
	Model	&$\bbeta$		&$\bz_{k}$&$\bz_{k0}$ \\
	\midrule
	$\rM_0$	&$\beta^{(c)}$		&1	&1	
	\\
	$\rM_t$	&$\bbeta^{(t)}$	&$\be_k$&$\be_k$
	\\
	$\rM_b$	&$(\beta^{(c)}, \beta^{(b)})^\T$				
	&$(1, f_{k})^\T$&$(1, 0)^\T$
	\\
  $\rM_{tb}$	&$(\bbeta^{(t)\T}, \beta^{(b)})^\T$				
	&$(\be_k^\T, f_{k})^\T$&$(\be_k^\T, 0)^\T$
	\\
	$\rM_h$	&$(\beta^{(c)}, \bbeta^{(h)\T})^\T$				
	&$(1, \bx^\T)^\T$	&$(1, \bx^\T)^\T$	
	\\
	$\rM_{ht}$	&$(\bbeta^{(h)\T}, \bbeta^{(t)\T})^\T$		
	&$(\bx^\T, \be_k^\T)^\T$&$(\bx^\T, \be_k^\T)^\T$
	\\
	$\rM_{hb}$	&$(\beta^{(c)}, \bbeta^{(h)\T}, \beta^{(b)})^\T$				
	&$(1, \bx^\T, f_{k})^\T$&$(1, \bx^\T, 0)^\T$
	\\
	$\rM_{htb}$	&$(\bbeta^{(h)\T}, \bbeta^{(t)\T},
	\beta^{(b)})^\T$				
	&$(\bx^\T, \be_k^\T, f_{k})^\T$&$(\bx^\T, \be_k^\T, 0)^\T$
	\\
	\bottomrule
	\end{tabular}
	
	\begin{tablenotes}
        	\footnotesize
        	\item $\be_k$ is the $K$-order vector whose $k$th
	component is 1 whereas all the other components are~0.
        	\item $f_{k} =1$ if $\sum_{j=0}^{k-1}d_{(j)}>0$ and $0$ otherwise.
   \end{tablenotes}
\end{threeparttable}     	
\end{table}

Since the effects of capture occasion and behavioral response
are both discrete,
 models $\rM_{0}$, $\rM_{t}$, $\rM_{b}$, and $\rM_{tb}$ are completely parametric
 \citep{otis1978statistical,chao2001overview},
 and EL is degenerate or inapplicable for these models.
Hereafter, we focus on $\rM_{h}$, $\rM_{ht}$, $\rM_{hb}$, and $\rM_{htb}$,
for which individual heterogeneity is vital.

\subsection{EL for general capture--recapture models}
\label{sec:emp-like-dis}

We begin by extending \cite{liu2017maximum}'s EL method to general capture--recapture models.
Let $\{(\bX_i^\T, D_{i1},\ldots, D_{iK}):i = 1,\dots,N\}$
be $N$ independent and identically distributed (i.i.d.) copies of $(\bX^\T, D_{(1)}, \ldots, D_{(K)})$, which is regarded
as an ideal infinite population.
Without loss of generality,
we suppose that the first $n$ individuals are captured
at least once
and that the observations are recorded as
$\{( \bx_i, d_{i1}, \dots, d_{iK}): i = 1, 2, \dots, n\}.$
We define $\bz_{ik}$ and $\bz_{ik0}$ in a similar way to $d_{ik}$.

Let $D = \sum_{k=1}^K D_{(k)}$ denote the
number of times that a generic individual is captured.
Then, $\alpha = \pr(D = 0)$
is the probability of an individual never being captured at all.
Based on the observations, the full likelihood is
\begin{align}
\label{full-likelihood}
\pr(n) \times \prod_{i=1}^n \pr(\bX = \bx_i \mid D>0)
\times \prod_{i=1}^n \pr(D_{(1)} = d_{i1}, \ldots,
D_{(K)} = d_{ik} \mid \bX= \bx_i, D>0).
\end{align}

Since $n$ has a binomial distribution $\rBi(N, 1 - \alpha)$,
then
\(
\pr(n) = \binom{N}{n} \alpha^{N - n}(1-\alpha)^n.
\)
It follows from $\bX \sim F(\bx)$ that the second term in~\eqref{full-likelihood} is
$$
\prod_{i=1}^n \frac{\pr(\bX = \bx_i)
\pr(D>0 \mid \bX = \bx_i)}{\pr(D>0)}
= \prod_{i=1}^n \frac{\rd F(\bx_i)
\{1 - \phi(\bx_i; \bbeta)\}}{1 - \alpha},
$$
where $\phi(\bx; \bbeta) = \pr(D=0 \mid \bX=\bx) =
\prod_{k=1}^K \{1 - g(\bz_{k0};\bbeta)\}$
is the probability of an individual never being captured
given its covariate $\bx$.
For model~\eqref{eq:cap-prob},
the third term in~\eqref{full-likelihood}~is
\begin{align*}
L_c(\bbeta) &:= \prod_{i=1}^n
\frac{\prod_{k=1}^K
\pr(  d_{ik} \mid  d_{i1}, \ldots,
  d_{ik-1},  \bx_i)}
{\pr(D>0 \mid \bX= \bx_i)}  =
\prod_{i=1}^n
\frac{\prod_{k=1}^K \{g(\bz_{ik}; \bbeta)\}^{d_{ik}}
\{1 - g(\bz_{ik}; \bbeta)\}^{1 - d_{ik}}}
{1 - \phi(\bx_i; \bbeta)}.
\end{align*}
In summary, the full likelihood~\eqref{full-likelihood}~is
\[
\binom{N}{n} \alpha^{N - n} \times
\prod_{i = 1}^n \rd F(\bx_i) \times
\prod_{i = 1}^n \prod_{k=1}^K \{g(\bz_{ik}; \bbeta)\}^{d_{ik}}
\{1 - g(\bz_{ik}; \bbeta)\}^{1 - d_{ik}}.
\]

In EL \citep{owen1988empirical, owen1990empirical},
we model the distribution of $\bX$ by a multinomial distribution
with support being the observations, i.e.\ $F(\bx)=\sum_{i=1}^ n p_iI(\bx_i\leq \bx)$, so that $\rd F(\bx_i) = p_i$.
Since $F(\bx)$ is a distribution function,
the feasible $p_i$'s should satisfy
\begin{align}\label{eq:constraint-pi-dis}
p_i\geq 0, \quad i =1,\dots,n, \quad
\sum_{i=1}^n p_i = 1, \quad
\sum_{i = 1}^n \{\phi(\bx_i; \bbeta) - \alpha\} p_i = 0,
\end{align}
where the last equation follows from
$\phi(\bx; \bbeta) = \pr(D=0 \mid \bX=\bx)$
and $\alpha = \pr(D = 0)$.
Substituting $p_i=\rd F(\bx_i)$ into the full likelihood
and taking logarithms give the log EL:
\begin{multline}
\label{eq:ell}
\widetilde\ell _e (N, \bbeta, \alpha, \{p_i\}) = \log \binom{N}{n} + (N - n) \log(\alpha)
+ \sum_{i = 1}^n \log(p_i) \\
 +
\sum_{i = 1}^n
\sum_{k = 1}^K [ d_{ik}\log\{g(\bz_{ik};\bbeta)\} + (1 - d_{ik})\log\{1 - g(\bz_{ik};\bbeta)\} ].
\end{multline}
Profiling out the $p_i$'s with the Lagrange multiplier method, we have
the profile log EL:
\begin{multline*}
\ell_e (N, \bbeta, \alpha) =
\log\binom{N}{n}+ (N - n) \log(\alpha)
- \sum_{i = 1}^n \log[1 + \xi\{\phi(\bx_i; \bbeta) - \alpha\}]
\\
+
\sum_{i = 1}^n
\sum_{k = 1}^K [ d_{ik}\log\{g(\bz_{ik};\bbeta)\} +
(1 - d_{ik})\log\{1 - g(\bz_{ik};\bbeta)\} ],
\end{multline*}
where $\xi = \xi(\bbeta, \alpha)$ satisfies
$
\sum_{i = 1}^n \frac{\phi(\bx_i; \bbeta) - \alpha}
{1 + \xi\{\phi(\bx_i; \bbeta) - \alpha\}} = 0.
$

Like \cite{liu2017maximum},
we define the maximum EL estimator of $(N, \bbeta, \alpha)$ as
$(\widehat N_e, \widehat\bbeta_e, \widehat\alpha_e) =
\arg\max \ell_e(N, \bbeta, \alpha).$
As an alternative, the Horvitz--Thompson type estimator of $N$
proposed by \cite{huggins1991some} is
$\widehat N_c = \sum_{i=1}^n
\{1 - \phi(\bx_i; \widehat\bbeta_c)\}^{-1}$,
where $\widehat\bbeta_c = \arg\max_{\bbeta}L_c(\bbeta)$ is the maximum CL estimator of $\bbeta$.

\subsection{Penalized empirical likelihood}
\label{sec:penalized-EL}

When the capture probability $(1 - \alpha)$ is small,
 the EL estimator $\widehat N_e$ can be unstable
and the corresponding interval estimates can be
extremely wide or even have infinite upper limits.
A possible reason is that the EL ratio function of $N$ increases too slowly or is even flat as $N$ increases.
Thus, we propose to penalize large values of $N$ in the (profile) log EL
function by adding a penalty. We define the penalized (profile) log EL function as:
\ba
\label{pel-discrete}
\widetilde\ell _p (N, \bbeta, \alpha, \{p_i\}) =
\widetilde\ell _e (N, \bbeta, \alpha, \{p_i\}) + C f(N),\;
\ell _p (N, \bbeta, \alpha)
 = \ell _e (N, \bbeta, \alpha) + C f(N),
\ea
where $f(N)$ is a non-increasing penalty function
and $C>0$ is a tuning parameter trading off the EL function
and the penalty term.
When $C=0$, the PEL method reduces to \cite{liu2017maximum}'s EL method.
Given $f(N)$ and $C$,
the proposed maximum PEL estimator is
$(\widehat N_p,\widehat \bbeta_p,\widehat \alpha_p ) =
\arg\max \ell_p(N, \bbeta, \alpha) $
and the proposed PEL ratio functions are
\ba
\label{pelr-discrete}
R_p(N, \bbeta, \alpha) = 2
\{\ell_p(\widehat N_p, \widehat\bbeta_p, \widehat\alpha_p) -
\ell_p(N, \bbeta, \alpha)\},\;
R_p'(N)
=
 \inf_{(\bbeta,\alpha)}R_p(N, \bbeta, \alpha).
\ea

The penalty $f(N)$ plays an important role in the PEL method.
To look for a reasonable function $f(N)$,
we recall \cite{chao1987estimating, chao1989estimating}'s
nonparametric estimator for $N$,
$\widetilde N_c = n + m_1^2/(2m_2)$,
where $m_1$ and $m_2$ are the numbers of individuals
captured once and twice, respectively.
Although negatively biased, this estimator is rather stable
and is widely used as a lower bound of $N$.
A desirable penalty $f(N)$ should shrink large $\widehat N_e$ toward Chao's lower bound $\widetilde N_c$ and make the log EL decrease quickly as $N$ increases.
Moreover, $f(N)$ should put less or no penalty on
small $\widehat N_e$ because the estimator itself is already stable.
These expectations motivated us to consider a quadratic penalty function,
$f(N) = -(N - \widetilde N_c)^2I(N> \widetilde N_c)$.
With an appropriate choice of $C$,
Theorem~\ref{thm:asy-pro} shows that the maximum PEL estimator
is asymptotically unbiased and asymptotically normal.

\begin{theorem}
\label{thm:asy-pro}
Let $(N_0, \bbeta_0, \alpha_0)$ with $\alpha_0\in(0,1)$ be
the true value of $(N, \bbeta, \alpha)$.
Suppose that the matrix $\bW$ defined in Equation~(1) of the supplementary material is positive definite.
When $f(N) = -(N - \widetilde N_c)^2I(N> \widetilde N_c)$
and $C = O_p(N_0^{-2})$, as $N_0\to\infty$, then:
 (a)
$\sqrt{N_0} \{\log(\widehat N_p/N_0),
(\widehat\bbeta_p-\bbeta_0)^\T,
\widehat\alpha_p - \alpha_0\}^\T
\convergeto \rN(\bzero,\bW^{-1})$,
where $\convergeto$ stands for convergence in distribution.
 (b)
$R_p(N_0, \bbeta_0, \alpha_0) \convergeto \chi^2_{2+s}$
and $R'_p(N_0) \convergeto \chi^2_{1}$,
where $s$ is the dimension of $\bbeta$ and $\chi^2_{df}$ is the chi-squared distribution
with $df$ degrees of freedom.

\end{theorem}

When $\bz_{k}$ is equal to $(1,\bx^\T)^\T$,
$\bW$ reduces to the matrix $W_s$
defined in Corollary~1 of \cite{liu2017maximum}.
Like the EL estimators of \cite{liu2017maximum},
 theoretically, the PEL estimators are equivalent to the CL estimators asymptotically,
and hence, they have the same limiting distributions.

\begin{proposition}
\label{cor:asy-equ}
Under the conditions in Theorem~\ref{thm:asy-pro}, as $ N _0 \to \infty$, then:
 (a)
$\widehat \bbeta_p - \widehat\bbeta_c = O_p(N_0^{-1}) $
and
$\widehat N_p - \widehat N_c = O_p(1)$.
 (b)
$\sqrt{ N _0}(\widehat \bbeta_p - \bbeta_0)
\convergeto \rN(\bzero, -\bV_{22}^{-1})$
and
$\sqrt{ N _0}(\widehat\bbeta_c - \bbeta_0)
\convergeto \rN(\bzero, -\bV_{22}^{-1})$.
 (c)
$ N _0^{-1/2} (\widehat N_p - N _0)
\convergeto \rN(0, \sigma^2)$
and
$ N _0^{-1/2} (\widehat N_c - N _0)
\convergeto \rN(0, \sigma^2)$,
where
$\sigma^2 = \varphi - 1 - \bV_{32}\bV_{22}^{-1}\bV_{23}$ and where $\varphi$ and the $\bV_{ij}$'s are defined in Section~1.1 of the supplementary material.
\end{proposition}

In practice, the performance of the PEL method depends on the tuning parameter $C$,
which may itself depend on the true value of $N$.
Like \cite{Wang2005penalized}, we recommend a data-adaptive value, $C = 2 m_2^2 /(n m_1^4)$,
which clearly satisfies the requirement $C = O_p(N_0^{-2})$ in Theorem~\ref{thm:asy-pro}.
From a Bayesian perspective,
adding the penalty $Cf(N)$ to the log EL is equivalent to
assuming a prior distribution:
$$
q(N) =
\frac{1}{2(\widetilde N_c-n)}I(n \leq N \leq \widetilde N_c)+
\frac{1}{\sqrt{2\pi}\widetilde\sigma_c}
\exp\left\{-\frac{(N - \widetilde N_c)^2}{2\widetilde\sigma_c^2}\right\}
I(N > \widetilde N_c),
$$
with $\widetilde\sigma_c^2 = n(\widetilde N_c - n)^2$,
for $N$.
When $N > \widetilde N_c$,
the prior $q(N)$ is a normal distribution
with mean $\widetilde N_c$ and variance $\widetilde\sigma^2_c$.
Otherwise, it reduces to a non-informative uniform distribution.
The EL estimator $\widehat N_e$ is hardly less than $\widetilde N_c$.
Therefore, the normal part of the prior takes effect in most cases.
The variance $\widetilde\sigma_c^2$ increases
and $C$ decreases as $(\widetilde N_c - n )$ increases.
This is reasonable since a larger gap between $\widetilde N_c$
and $n$ implies that more individuals are not sampled and there is more uncertainty in the data.

\section{EM algorithms}
\label{sec:alg}

The main numerical task in the proposed PEL approach to abundance estimation is
to maximize the penalized log EL function
$\widetilde\ell_p(N, \bbeta, \alpha, \{p_i\})$ or
the penalized profile log EL $\ell_p(N, \bbeta, \alpha)$.
In the implementation of the EL method,
\cite{liu2017maximum} proposed using 
optimization functions.
However, the calculated results may depend on the choice of initial values and be unreliable.
Considering that the EM algorithm \citep{dempster1977maximum} is well known for its stability, we develop EM algorithms for the PEL method
to improve the numerical performance of the EL abundance estimation method.

\subsection{Preparation}

As we assumed in Section~\ref{sec:emp-like-dis}, only
the first $n$ individuals are observed.
Given the population size $N$,
we use $\bx_j^*$'s to denote the covariates of the other $N-n$ individuals. The $\bx_j^*$'s are independent of each other
and are also independent of $\bx_i$ ($i=1, \dots, n$).
They have a common distribution $F_{\bX}$ and
serve as missing data in the subsequent EM algorithm.
We regard $\bf O^* \cup O$ as complete data, where
${\bf O^*} = \{ \bx_j^*: j = n+1,\dots, N\}$ and
\begin{multline*}
{\bf O} = \{(\bx_i, d_{i1}, \ldots, d_{iK}) : 1\leq i\leq n\}
\cup
\{ (d_{i1}, \ldots, d_{iK}) : d_{ik} = 0, 1\leq k\leq K, n+1\leq i\leq N\}.
\end{multline*}

Let $\btheta = (\bbeta^\T, \alpha)^\T$ and
$\bpsi = (\btheta^\T, p_1, \dots, p_n)^\T$.
In this case, the complete-data likelihood~is
\begin{multline*}
\prod_{i = 1}^n \left[\left \{
\prod_{k = 1}^K \pr(  d_{ik} \mid   \bx_i,
 d_{i1},\dots,   d_{ik-1}) \right \}
\pr(\bX_i= \bx_i) \right]
\times
\prod_{j = n+1}^N
\{\phi(\bx_j^*; \bbeta)\pr(\bX_j = \bx_j^*)\},
\end{multline*}
and the corresponding log likelihood of $\bpsi$ is
\begin{multline*}
\ell(\bpsi) =
\sum_{i = 1}^n
\sum_{k = 1}^K [
d_{ik}\log\{g(\bz_{ik};\bbeta)\} +
(1- d_{ik})\log\{1 - g(\bz_{ik};\bbeta)\}] +
\sum_{i = 1}^n\log(p_i) \\
+ \sum_{i = 1}^n \sum_{j = n+1}^N
\left[ I(\bX_j = \bx_i)
\log\{\phi(\bx_{i};\bbeta)p_i\}
\right],
\end{multline*}
where the last term is obtained from the definition of the $p_i$'s and since the $\bX_j$'s take values from $\{\bx_1,\dots,\bx_n\}$.

As in \cite{dempster1977maximum},
the standard EM algorithm consists of a sequence of iterations.
Each iteration involves two steps: an E-step and an M-step.
Suppose that $r$ iterations ($r=0,1,\dots$) have been finished.
In the $(r+1)$th iteration, we need to calculate the expectation of the above log likelihood
$\ell(\bpsi)$ conditioned on the observed data $\bf O$
and given $\bpsi = \bpsi^{(r)}$.
For $j = n+1,\dots, N$,
it follows from $\bX_j\sim F_{\bX}$ that
\[
\e
\left\{I(\bX_j = \bx_i) \mid
{\bf O}, \bpsi = \bpsi^{(r)}\right\}
= \pr(\bX_j = \bx_i \mid D_{j1} = \cdots = D_{jK} = 0)
= \frac{\phi(\bx_i; \bbeta^{(r)})p_i^{(r)}}
{\alpha^{(r)}},
\]
where $\alpha^{(r)} = \sum_{i = 1}^n \phi(\bx_i; \bbeta^{(r)})p_i^{(r)}$.
Thus, the conditional expectation of
$\ell(\bpsi)$ can be written as
$\mQ (\bpsi \mid \bpsi^{(r)})
= \ell_1(\bbeta) + \ell_2(p_1,\ldots,p_n)$, where
\begin{align*}
\ell_1(\bbeta) = \sum_{i = 1}^n &
\sum_{k = 1}^K [
d_{ik}\log\{g(\bz_{ik};\bbeta)\} +
(1 - d_{ik})\log\{1 - g(\bz_{ik};\bbeta)\} +
w_i^{(r)} \log\{1 - g(\bz_{ik0};\bbeta)\}], \\
\ell_2(p_1,\ldots,p_n) &= \sum_{i = 1}^n (w_i^{(r)} + 1) \log(p_i), \quad
w_i^{(r)} = (N - n)
\phi(\bx_i;\bbeta^{(r)}) p_i^{(r)}/\alpha^{(r)}.
\end{align*}
This completes the E-step of the EM algorithm for the $(r+1)$th iteration.

Now, the M-step of the EM algorithm is undertaken
by choosing $\bpsi = \bpsi^{(r+1)}$,
which maximizes $\mQ(\bpsi \mid \bpsi^{(r)})$
with respect to $\bpsi$ under the constraint~\eqref{eq:constraint-pi-dis}.
Since $\mQ(\bpsi \mid \bpsi^{(r)})$ does not involve $\alpha$,
we propose to update $\alpha^{(r)}$ as
$\alpha^{(r+1)} = \sum_{i = 1}^n \phi(\bx_i; \bbeta^{(r+1)})p_i^{(r+1)}$.
Here, the $p^{(r+1)}_i$'s are obtained by maximizing $\ell_2(p_1,\ldots,p_n)$
such that
$p_i\geq 0$ ($i =1,\dots,n)$ and
$\sum_{i=1}^n p_i = 1.$
The maximizer is
$p_i^{(r+1)} = (w_i^{(r)} + 1)/N$, $i=1,\dots,n.$
Then,
$\bbeta^{(r+1)}$ is obtained by maximizing $\ell_1(\bbeta)$,
which can be solved by the standard Newton--Raphson method or
by fitting a binomial regression model.

\subsection{Algorithm}
\label{sec:proc-EM}

Based on the preceding analysis,
we propose using the following EM algorithm
to maximize the penalized log EL function
$\widetilde\ell_p(N, \bbeta, \alpha, \{p_i\})$ for a given $N$.

\begin{description}
\item[Step 0.]
Set $\bbeta^{(0)} = \bf 0$, $p_i^{(0)} = 1/n$
for $i = 1, \dots, n$,
$\alpha^{(0)} = \sum_{i = 1}^n \phi(\bx_i; \bbeta^{(0)})p_i^{(0)}$,
and the iteration number $r = 0$.
\item[Step 1.]
Calculate $w_i^{(r)} = (N - n)
\phi(\bx_i;\bbeta^{(r)}) p_i^{(r)}/\alpha^{(r)}$.
Update
$\bbeta^{(r)}$ to $\bbeta^{(r+1)}$
by fitting a binomial regression model with a logistic link function
to the observations
$\{(y_{ik}, \bz_{ik}^\T): i=1,\dots,n; k = 1,\dots,K\}$ and
$\{(0, \bz_{ik0}^\T): i=1,\dots,n; k = 1,\dots,K\}$
with weights 1's and $w_i^{(r)}$'s, respectively.

\item[Step 2.]
Update $p_i^{(r)}$ to $p_i^{(r+1)} = (w_i^{(r)} + 1)/N$ for $i=1,\dots,n$,
and calculate $\alpha^{(r+1)} = \sum_{i = 1}^n \phi(\bx_i; \bbeta^{(r+1)})p_i^{(r+1)}$.

\item[Step 3.]
Set $r =r +1$ and
repeat steps 1 and 2 until the increment of the penalized log EL in Equation~\eqref{eq:ell}
after an iteration is no greater than a tolerance, say, $10^{-5}$.
\end{description}

\begin{remark}\label{mak:em-el}
The parameter $N$ is fixed in the above algorithm.
To calculate the maximum PEL estimator
$(\widehat N_p,\widehat \bbeta_p,\widehat \alpha_p)$,
we have two alternative methods.
One is to directly maximize the profile function
$\max_{(\bbeta, \alpha, \{p_i\})} \widetilde\ell_p(N, \bbeta, \alpha, \{p_i\})$
with respect to $N$.
The other is to use $w_i^{(r)} = (N^{(r)} - n)
\phi(\bx_i;\bbeta^{(r)}) p_i^{(r)}/\alpha^{(r)}$ and
$p_i^{(r+1)} = (w_i^{(r)} + 1)/N^{(r)}$ in steps 1 and 2, respectively,
and add the following maximization step after step~2:
\begin{description}
\item[Step 2$'$.]
Update $N^{(r)}$ to $N^{(r+1)}$, the maximizer of
\(
\log\binom{N}{n} + (N - n) \log(\alpha^{(r+1)}) + C f(N).
\)
\end{description}
The unpenalized EL estimator $(\widehat N_e,\widehat \bbeta_e,\widehat \alpha_e)$
can be calculated with the same algorithm after setting $C=0$.
\end{remark}

We have integrated the above EM algorithm into the {\tt R} package {\tt Abun},
where steps 1 and 2$'$ are, respectively, implemented via {\tt R} functions {\tt glm} and {\tt optimize}.
The use of standard GLM classes makes
the EM algorithm very reliable and flexible.
Besides this advantage, the proposed EM algorithm
inherits many appealing properties of the classical EM algorithm.

\begin{theorem}\label{thm:em}
With discrete-time capture--recapture models,
the EM algorithm proposed for the PEL method has following properties:
 (a)  When $N$ is fixed, the penalized log EL is nondecreasing
after each EM iteration.
 (b)
When $N$ is unknown, the penalized log EL is nondecreasing after each EM iteration.
 (c)
When $N$ is unknown, the sequence of EM iterations $(N^{(r)}, \bbeta^{(r)}, \alpha^{(r)})$
converges to a local maximum PEL estimator
$(\widehat N_p, \widehat\bbeta_p, \widehat\alpha_p)$.

\end{theorem}

\begin{remark}\label{mak:em-cl}
The proposed EM algorithm is applicable to the CL method
if $N^{(r+1)}$ is set to $n/(1-\alpha^{(r+1)})$ in step~2$'$ and
the iteration stops when the log CL $\log\{L_c(\bbeta)\}$ converges in step~3.
A justification for setting $N^{(r+1)} = n/(1-\alpha^{(r+1)})$ is that
if we regard $N$ as a random variable,
it is reasonable to assume that $N$ given $n$ follows a negative binomial distribution,
which implies that $\e(N \mid n) = n/(1 - \alpha)$.
\end{remark}

\begin{proposition}\label{cor:em}
Considering Remark~\ref{mak:em-cl},
the sequence of EM iterations $\{(N^{(r)}, \bbeta^{(r)}): r=1,2, \dots\}$
converges to a local maximum CL estimator
$(\widehat N_c, \widehat\bbeta_c)$.
\end{proposition}

Although the proposed EM algorithm is designed for capture--recapture models with individual heterogeneity, if we set the coefficient of $\bx$ to $\bzero$ in step~2, then
it is also applicable to models without individual heterogeneity,
such as $\rM_0$, $\rM_t$, $\rM_b$, and $\rM_{tb}$.

\section{Simulation study}
\label{sec:sim}
In this section, we carry out simulations to investigate the finite-sample performance of
the PEL methods and the proposed EM algorithms for discrete-time capture--recapture models.
For point estimation of $N$,
we study the maximum PEL estimator $\widehat N_p$,
the maximum CL estimator $\widehat N_c$, and the EL estimator $\widehat N_e$.
For interval estimation, we compare
\bit
\item[(1)]
the PEL ratio confidence interval $\mI_p = \{N: R'_p(N) \leq \chi^2_1(1-a)\}$,
\item[(2)]
the EL ratio confidence interval $\mI_e= \{N: R'_e(N) \leq \chi^2_1(1-a)\}$,
and
\item[(3)]
the Wald-type confidence interval $\mI_c= \{N: (\widehat N_c - N)^2/(\widehat N_c \widehat\sigma_c^2)
\leq \chi^2_1(1-a) \}$,
\eit
where $\chi^2_1(1-a)$ is the $(1-a)$th quantile of
$\chi^2_1$ and
$\widehat \sigma^2_c$ is a consistent estimate of $\sigma^2$.
Theorem~\ref{thm:asy-pro} 
together with
Proposition~\ref{cor:asy-equ} 
guarantees that these confidence intervals all
have a coverage probability of $(1-a)$ asymptotically.

We consider four interesting questions:
\bit
\item[]
\textit{Question 1.} Is the EM algorithm more reliable than the standard optimization algorithm?
\item[]
\textit{Question 2.} How does individual behavior effect
the maximum EL estimator $\widehat N_e$?
\item[]
\textit{Question 3.}
Is the maximum PEL estimator $\widehat N_p$ more stable than its competitors?
\item[]
\textit{Question 4.}
Is the PEL ratio confidence interval $\mI_p$ superior to its competitors?

\eit
To answer these questions, we generate data for the following three scenarios:
\bit

\item[(A)]
Let $\bX = (X_1, X_2)^\T$, where $X_1\sim\rN(0,1)$ and $X_2\sim\rBi(1,0.5)$.
We consider model $\rM_h$ with $\bbeta = (0.1, -2.5, -0.15)^\T$.

\item[(B)]
The settings are the same as (A) except that we use model $\rM_{hb}$ with
 $\bbeta _0 =$ \linebreak $ (0.1, -2.5, -0.15, 0.8)^\T$.

\item[(C)] The settings are the same as (B) except that
$\bbeta_0 = (0.1, -2.5, -0.15, -0.8)^\T$.

\eit
In each scenario, we set the population size to $N_0=200$ or 400
and the number of capture occasions to $K=2$ or 6,
where the capture probability is about 63\% or 79\%.
All our simulation results were calculated for 5000 samples.

\begin{figure}[bt]
\begin{center}
\includegraphics[width=0.32\textwidth]{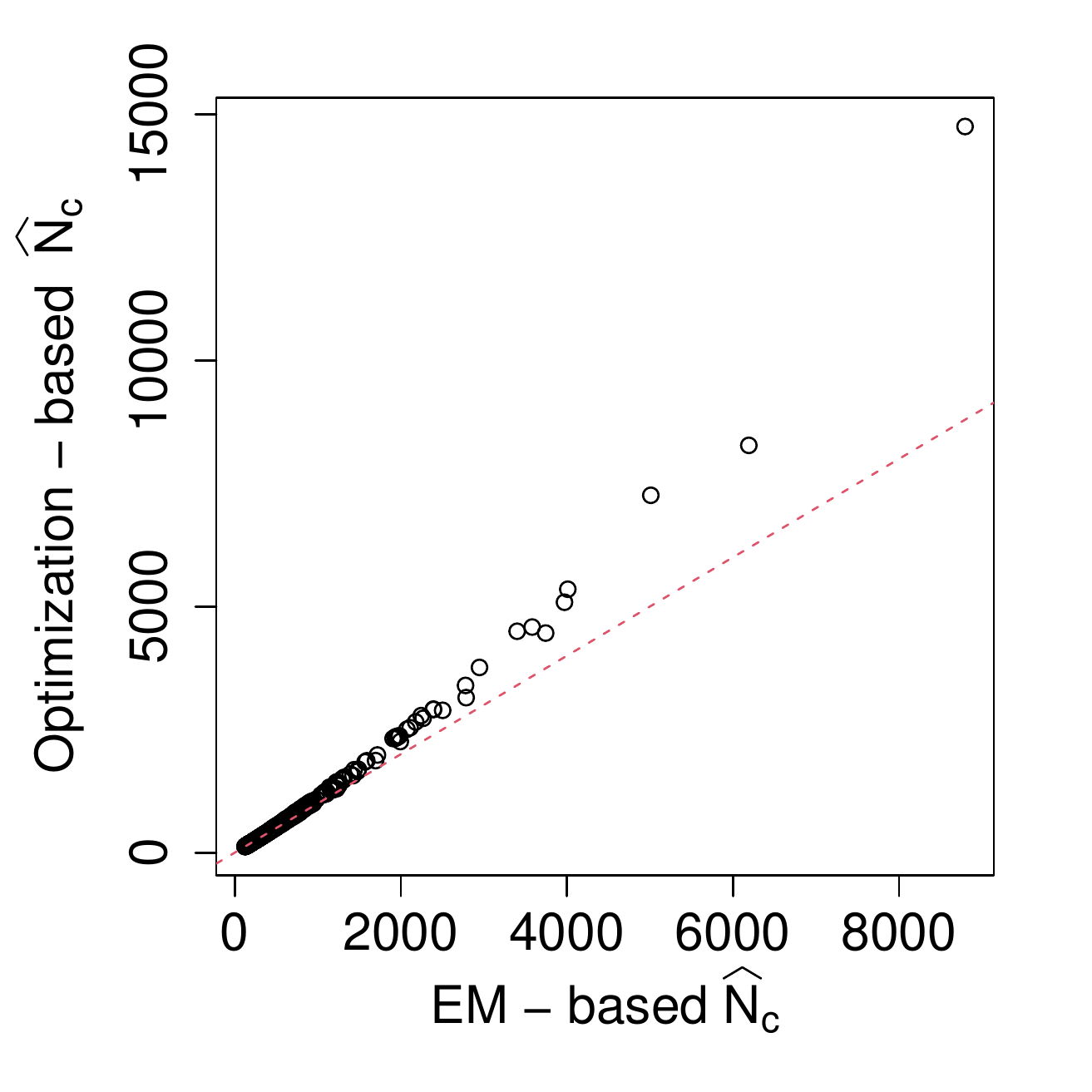}
\includegraphics[width=0.32\textwidth]{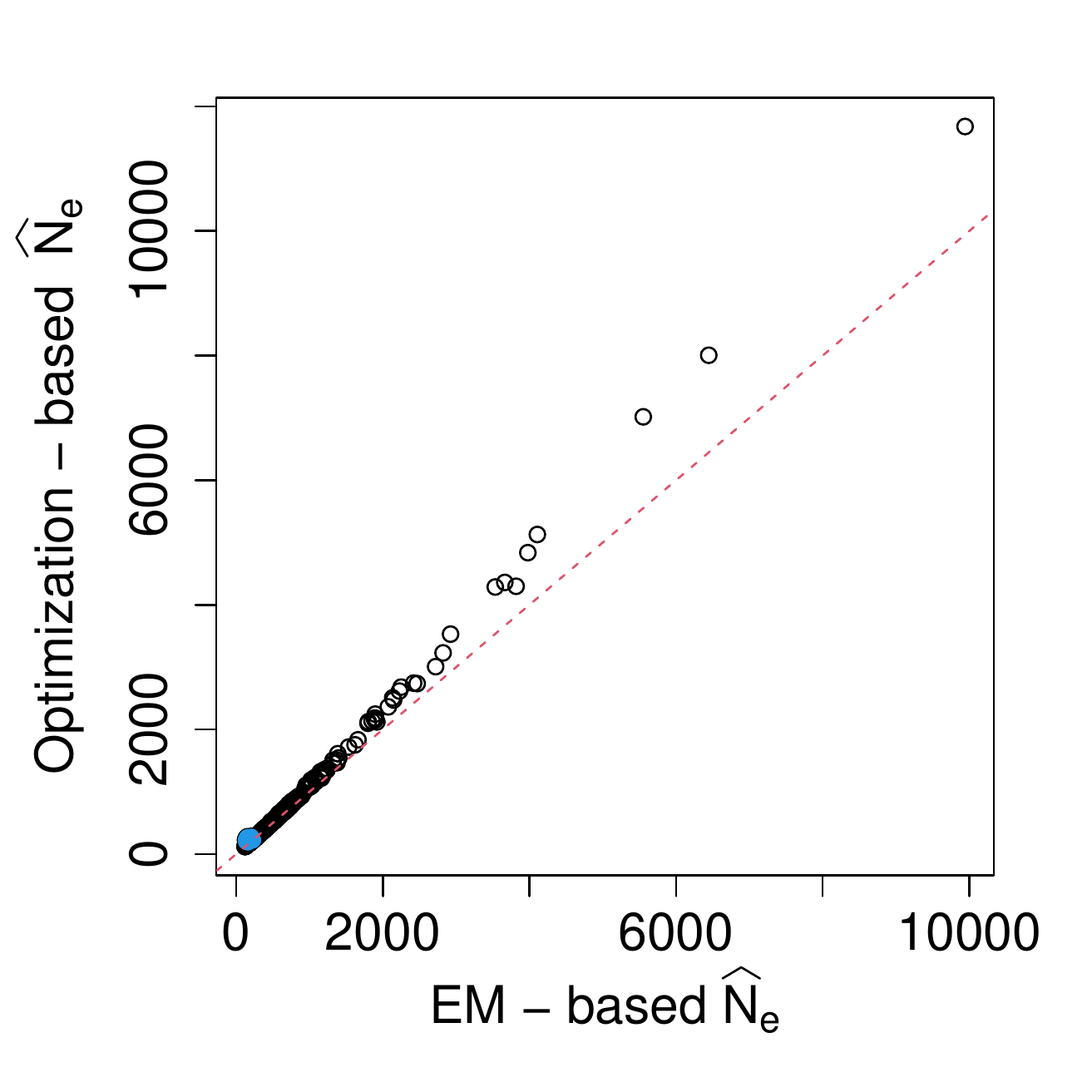}
\includegraphics[width=0.32\textwidth]{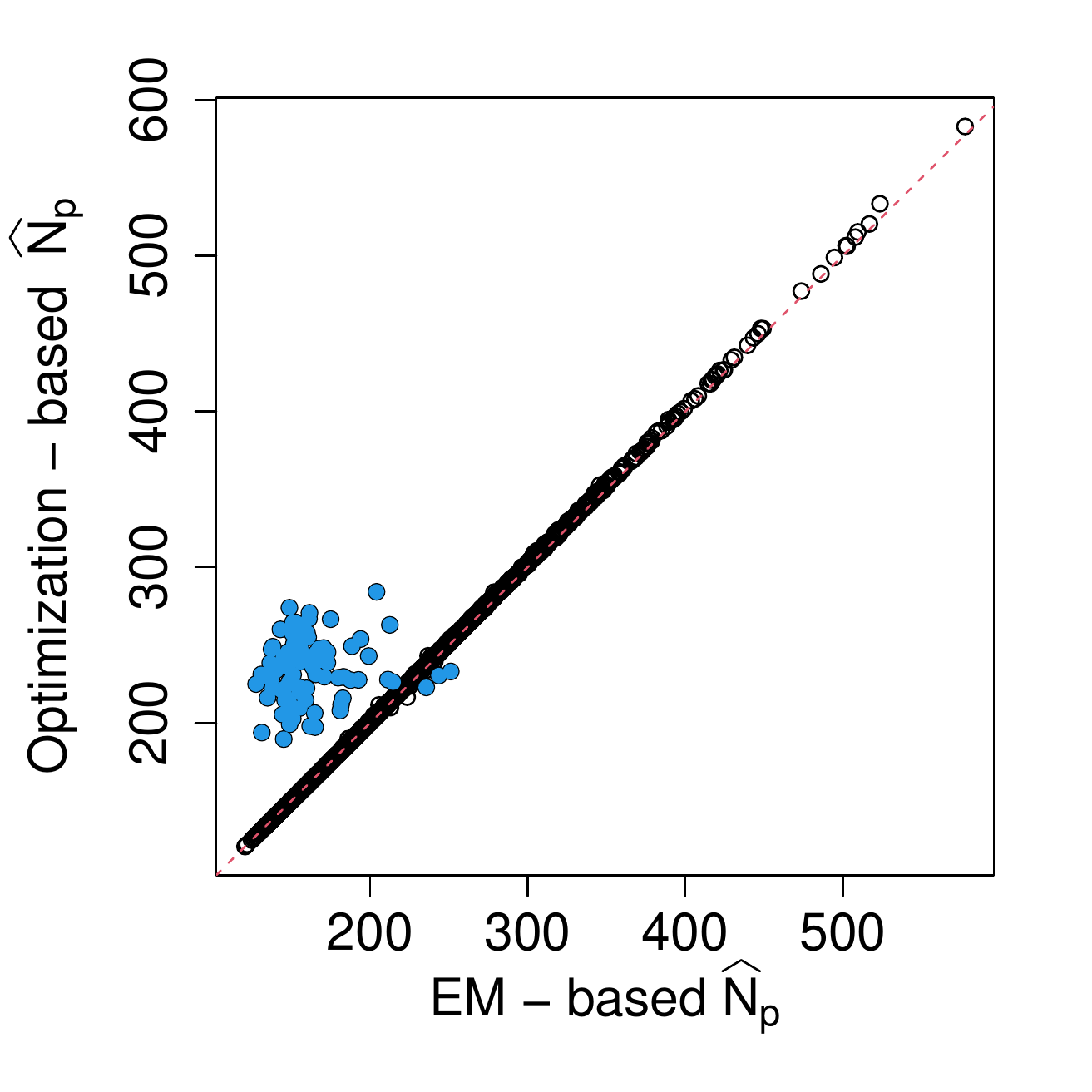}
\end{center}
 \caption{Scatter plots of the EM-based versus optimization-based estimates.
 The solid blue dots show  where
the EM-based log likelihoods are greater than
 the optimization-based log likelihoods by 0.01 or more.}
\label{fig:scatterplot}	
\end{figure}

\paragraph{Answer to question~1.}
We calculate the three point estimators of $N$ with both the proposed EM algorithm
and the optimization algorithm of \cite{liu2017maximum}.
Figure~\ref{fig:scatterplot} 
displays the scatter plots of the EM-based versus optimization-based point estimates
when data were generated for Scenario A with $N_0=200$ and $K = 2$.
For all three estimators, $\widehat N_c$, $\widehat N_e$, and
$\widehat N_p$, the optimization algorithm
tends to produce larger estimates than the EM algorithm does, especially when
the EM-based estimates are close to or greater than $N_0=200$.
To some extent, this implies that the optimization algorithm is
less robust than the EM algorithm.
Moreover, there are quite a few cases
where the log likelihoods based on the optimization algorithm
are less than those based on the EM algorithm by 0.01 or more.
This indicates that in these cases, the optimization algorithm
might not find the maximum likelihood estimates
even when the estimates themselves are not large.
It also suggests that the proposed EM algorithm is
more reliable than the optimization algorithm.
In the following, unless stated otherwise,
all estimates are calculated by
the 
EM algorithm. 

\begin{figure}[bt!]
\begin{center}
\makebox{
\includegraphics[width=0.32\textwidth]{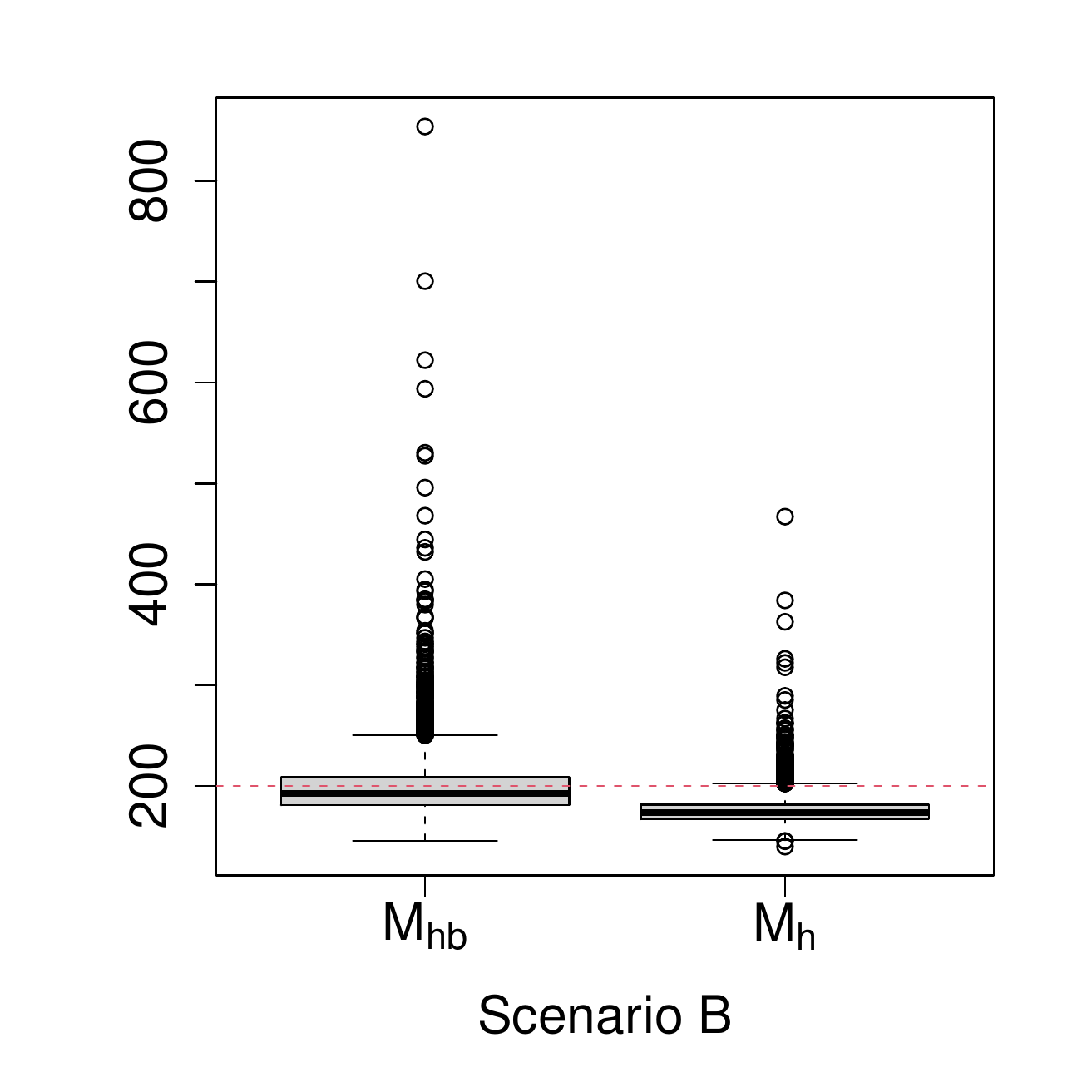}
\includegraphics[width=0.32\textwidth]{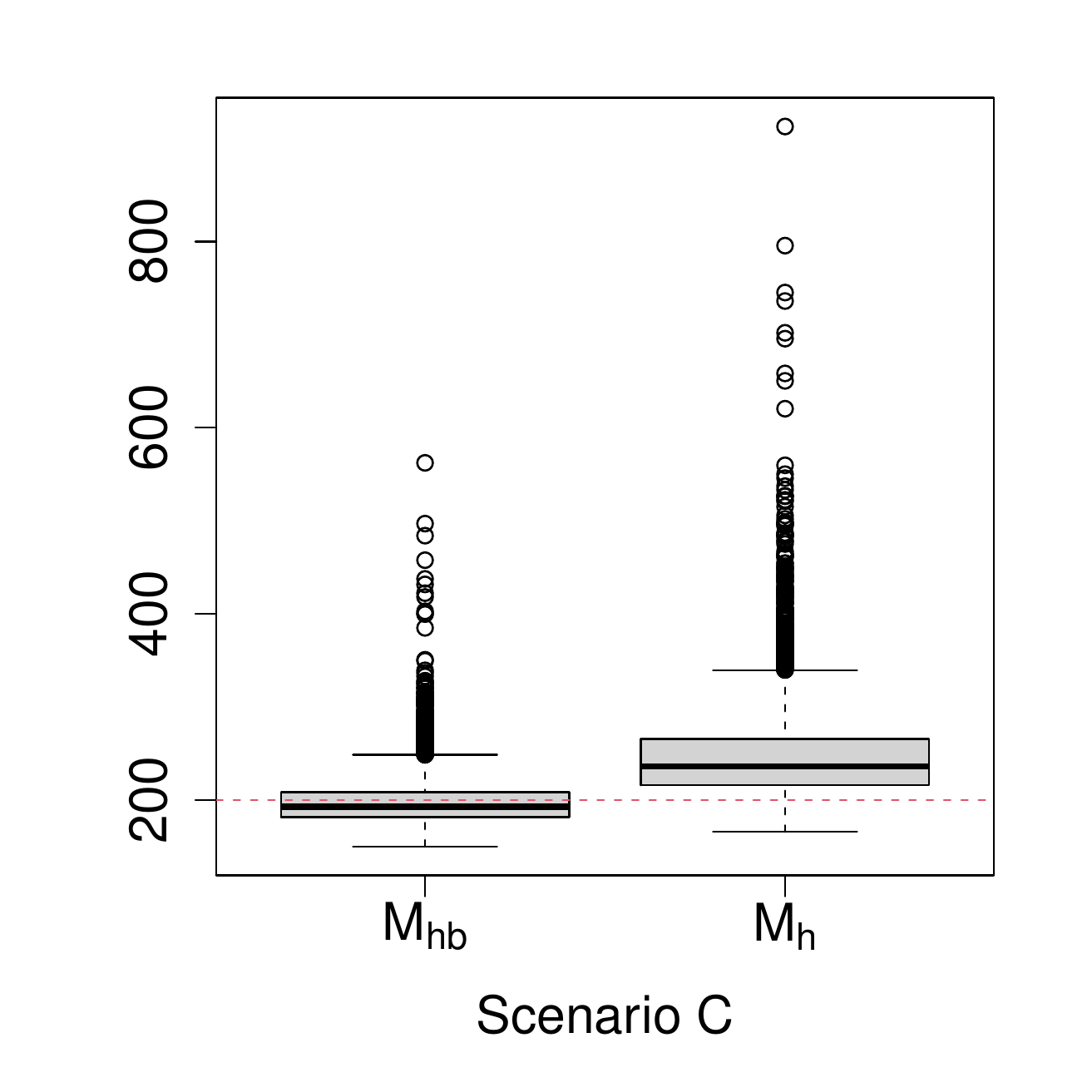}
}
\end{center}
 \caption{Box plots of $\widehat N_e$ for models $\rM_{h}$ and $\rM_{hb}$
 for scenarios B (left) and C (right).}
\label{fig:boxplot-N}	
\end{figure}

\paragraph{Answer to question~2.}
To investigate the impact of individual behavior,
we consider the EL estimator $\widehat N_e$
for the $\rM_{h}$ and $\rM_{hb}$ models.
Figure~\ref{fig:boxplot-N} displays
the box plots of $\widehat N_e$ when data were generated for scenarios B and C with $N_0 = 200$ and $K = 6$.
Since model $\rM_{hb}$ is satisfied in both scenarios B and C,
we expect that the EL estimator
for $\rM_{h}$ may perform less well,
whereas for $\rM_{hb}$, it should do better.
From Figure~\ref{fig:boxplot-N}, we see that the $\rM_{hb}$-based $\widehat N_e$
is nearly unbiased, as expected.
However, the $\rM_{h}$-based $\widehat N_e$, which ignores the behavior effect,
produces obvious underestimates for scenario B and obvious overestimates for scenario C.
We also see that $\widehat N_e$ has increasingly stable performance
as the capture probability increases (63\% in scenario B and 79\%
in scenario C).
Similar simulations were conducted for the PEL estimator $\widehat N_p$, and the behavior response had the same influence on~$\widehat N_p$.

\paragraph{Answer to question~3.}
We use the root mean square error (RMSE) to evaluate
the stability of a point estimator for the population size $N$.
Table~\ref{tab:sim} presents
the true value $N_0$ and the RMSEs of
the PEL estimator $\widehat N_p$ and its competitors:
the CL estimator $\widehat N_c$ and
the EL estimator $\widehat N_e$.
We also include the RMSE of the CL estimator $\widehat N_v$
calculated by the {\tt R} package {\tt VGAM}.

\begin{table}[bt!]
\centering
\caption{Root mean square errors (RMSEs) of point estimates
and coverage probabilities of interval estimates at the 95\% level.}
\label{tab:sim}
	\begin{tabular}{ C{1cm}C{1cm}C{1cm}C{1cm}C{1cm}
	C{0.5cm}C{1cm}C{1cm}C{1cm}C{1cm}C{1cm} }
	\toprule
&&\multicolumn{4}{c}{RMSE}&&
\multicolumn{4}{c}{Level: 95\%}
\\
$K$ & $N_0$ & $\widehat N_v$ & $\widehat N_c$ & $\widehat N_e$ &
$\widehat N_{p}$ &  & $\mI_v$ & $\mI_c$ & $\mI_e$ & $\mI_{p}$
\\
\midrule
\multicolumn{11}{c}{Scenario A}\\
2	  & 200  & 366 & 271 & 273 & 50 &  & 87.52 & 87.72 & 92.66  & 93.22
\\
	  & 400  & 191 & 167 & 156 & 88 &  & 90.18 & 90.26 & 93.74 & 93.92
\\
6	  & 200  & 32 & 30 & 27 & 24 &  & 87.14 & 87.44  & 90.94 & 90.76
\\
	  & 400  & 44 & 41 & 38 & 36 &  & 88.62 & 88.66 & 91.24 & 91.26
\\
\multicolumn{11}{c}{Scenario B}\\
2	  & 200  & 70\,302 & 10\,271 & 726 & 44 &  & 87.92 & 88.00 & 92.70 & 92.92
\\
	  & 400  & 626 & 381 & 394 & 86 &  & 89.64 & 89.54 & 93.58 & 94.24
\\
6	  & 200  & 38 & 36 & 32 & 24 &  & 87.44 & 87.68 & 91.72 & 90.22
\\
	  & 400  & 49 & 46 & 43 & 37 &  & 89.52 & 89.38 & 91.80 & 91.70
\\
\multicolumn{11}{c}{Scenario C}\\
2	  & 200  & 563 & 383 & 386 & 72 &  & 87.12 & 87.12 & 92.88 & 93.84
\\
	  & 400  & 215 & 192 & 176 & 120 &  & 89.28 & 89.26 & 93.36 & 93.64
\\
6	  & 200  & 32 & 31 & 28 & 26 &  & 87.12 & 87.50 & 91.58 & 91.56
\\
	  & 400  & 44 & 41 & 39 & 38 &  & 89.48 & 89.48 & 92.16 & 92.12
\\
\bottomrule
	\end{tabular}
\end{table}

In all scenarios, the PEL estimator $\widehat N_p$ always
has the smallest RMSEs, indicating that it has the most stable
performance of these four estimators.
In particular, when $K=2$ and $N_0=200$,
the RMSEs of $\widehat N_c$, $\widehat N_v$, and $\widehat N_e$
are even greater than $N_0$ itself, which is undesirable.
In contrast, the PEL estimator has significantly smaller RMSEs, 80\% or more lower, which is very surprising.
When $K=6$, all four estimators are very stable although
the PEL estimator still performs best.
In summary, we have an affirmative answer to question~3, namely
the maximum PEL estimator $\widehat N_p$ is, indeed, more stable than its competitors.

\paragraph{Answer to question~4.}
Parallel to the four point estimators,
we compare the finite-sample performances of
four interval estimators
($\mI_c$, $\mI_e$, $\mI_p$, and $\mI_v$)
where $\mI_v$ is the Wald-type confidence interval
calculated by the {\tt R} package {\tt VGAM}.
We report the coverage probabilities of the four interval estimators
at the 95\% confidence level in Table~\ref{tab:sim}.
The PEL interval $\mI_{p}$ always has the same coverage probabilities
as the EL interval $\mI_{e}$,
and both have much better coverage accuracy than
the two CL intervals $\mI_c$ and $\mI_v$.
The increase in coverage for the PEL and EL intervals is at least 2\% and can be as high as 6\%.
For example, when $K=2$ and $N_0=200$ for scenario C, the increases in coverage are 6.72\% and 5.76\%, respectively.

\begin{figure}[bt!]
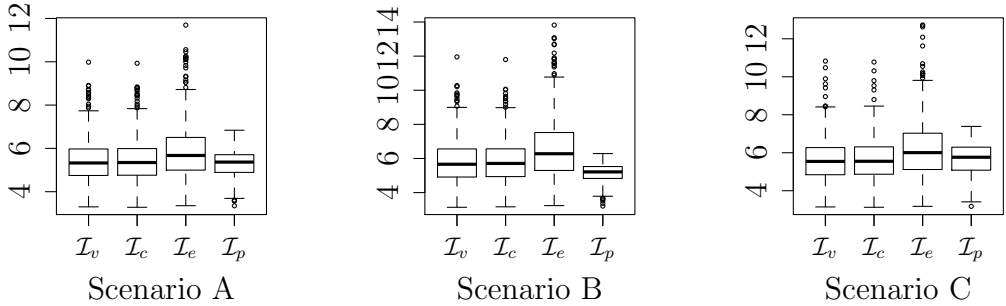

\begin{center}
\input{ciA200-2}
\input{ciB200-2}
\begin{tikzpicture}[x=0.7pt,y=0.7pt]
\definecolor{fillColor}{RGB}{255,255,255}
\path[use as bounding box,fill=fillColor,fill opacity=0.00] (0,0) rectangle (187.90,216.81);
\begin{scope}
\path[clip] ( 49.20, 61.20) rectangle (162.70,167.61);
\definecolor{drawColor}{RGB}{0,0,0}

\path[draw=drawColor,line width= 1.2pt,line join=round] ( 56.03, 89.94) -- ( 77.05, 89.94);

\path[draw=drawColor,line width= 0.4pt,dash pattern=on 4pt off 4pt ,line join=round,line cap=round] ( 66.54, 65.33) -- ( 66.54, 82.75);

\path[draw=drawColor,line width= 0.4pt,dash pattern=on 4pt off 4pt ,line join=round,line cap=round] ( 66.54,119.42) -- ( 66.54, 97.42);

\path[draw=drawColor,line width= 0.4pt,line join=round,line cap=round] ( 61.29, 65.33) -- ( 71.80, 65.33);

\path[draw=drawColor,line width= 0.4pt,line join=round,line cap=round] ( 61.29,119.42) -- ( 71.80,119.42);

\path[draw=drawColor,line width= 0.4pt,line join=round,line cap=round] ( 56.03, 82.75) --
	( 77.05, 82.75) --
	( 77.05, 97.42) --
	( 56.03, 97.42) --
	( 56.03, 82.75);

\path[draw=drawColor,line width= 0.4pt,line join=round,line cap=round] ( 66.54,129.71) circle (  1.12);

\path[draw=drawColor,line width= 0.4pt,line join=round,line cap=round] ( 66.54,140.61) circle (  1.12);

\path[draw=drawColor,line width= 0.4pt,line join=round,line cap=round] ( 66.54,119.82) circle (  1.12);

\path[draw=drawColor,line width= 0.4pt,line join=round,line cap=round] ( 66.54,144.18) circle (  1.12);

\path[draw=drawColor,line width= 0.4pt,line join=round,line cap=round] ( 66.54,119.49) circle (  1.12);

\path[draw=drawColor,line width= 0.4pt,line join=round,line cap=round] ( 66.54,119.83) circle (  1.12);

\path[draw=drawColor,line width= 0.4pt,line join=round,line cap=round] ( 66.54,125.01) circle (  1.12);

\path[draw=drawColor,line width= 0.4pt,line join=round,line cap=round] ( 66.54,134.64) circle (  1.12);

\path[draw=drawColor,line width= 0.4pt,line join=round,line cap=round] ( 66.54,131.83) circle (  1.12);

\path[draw=drawColor,line width= 1.2pt,line join=round] ( 82.30, 90.01) -- (103.32, 90.01);

\path[draw=drawColor,line width= 0.4pt,dash pattern=on 4pt off 4pt ,line join=round,line cap=round] ( 92.81, 65.14) -- ( 92.81, 83.01);

\path[draw=drawColor,line width= 0.4pt,dash pattern=on 4pt off 4pt ,line join=round,line cap=round] ( 92.81,119.87) -- ( 92.81, 97.82);

\path[draw=drawColor,line width= 0.4pt,line join=round,line cap=round] ( 87.56, 65.14) -- ( 98.07, 65.14);

\path[draw=drawColor,line width= 0.4pt,line join=round,line cap=round] ( 87.56,119.87) -- ( 98.07,119.87);

\path[draw=drawColor,line width= 0.4pt,line join=round,line cap=round] ( 82.30, 83.01) --
	(103.32, 83.01) --
	(103.32, 97.82) --
	( 82.30, 97.82) --
	( 82.30, 83.01);

\path[draw=drawColor,line width= 0.4pt,line join=round,line cap=round] ( 92.81,128.57) circle (  1.12);

\path[draw=drawColor,line width= 0.4pt,line join=round,line cap=round] ( 92.81,138.91) circle (  1.12);

\path[draw=drawColor,line width= 0.4pt,line join=round,line cap=round] ( 92.81,143.68) circle (  1.12);

\path[draw=drawColor,line width= 0.4pt,line join=round,line cap=round] ( 92.81,123.40) circle (  1.12);

\path[draw=drawColor,line width= 0.4pt,line join=round,line cap=round] ( 92.81,133.61) circle (  1.12);

\path[draw=drawColor,line width= 0.4pt,line join=round,line cap=round] ( 92.81,130.89) circle (  1.12);

\path[draw=drawColor,line width= 1.2pt,line join=round] (108.58, 94.69) -- (129.60, 94.69);

\path[draw=drawColor,line width= 0.4pt,dash pattern=on 4pt off 4pt ,line join=round,line cap=round] (119.09, 65.64) -- (119.09, 85.56);

\path[draw=drawColor,line width= 0.4pt,dash pattern=on 4pt off 4pt ,line join=round,line cap=round] (119.09,133.77) -- (119.09,105.20);

\path[draw=drawColor,line width= 0.4pt,line join=round,line cap=round] (113.83, 65.64) -- (124.34, 65.64);

\path[draw=drawColor,line width= 0.4pt,line join=round,line cap=round] (113.83,133.77) -- (124.34,133.77);

\path[draw=drawColor,line width= 0.4pt,line join=round,line cap=round] (108.58, 85.56) --
	(129.60, 85.56) --
	(129.60,105.20) --
	(108.58,105.20) --
	(108.58, 85.56);

\path[draw=drawColor,line width= 0.4pt,line join=round,line cap=round] (119.09,135.02) circle (  1.12);

\path[draw=drawColor,line width= 0.4pt,line join=round,line cap=round] (119.09,152.36) circle (  1.12);

\path[draw=drawColor,line width= 0.4pt,line join=round,line cap=round] (119.09,163.67) circle (  1.12);

\path[draw=drawColor,line width= 0.4pt,line join=round,line cap=round] (119.09,138.31) circle (  1.12);

\path[draw=drawColor,line width= 0.4pt,line join=round,line cap=round] (119.09,162.72) circle (  1.12);

\path[draw=drawColor,line width= 0.4pt,line join=round,line cap=round] (119.09,141.35) circle (  1.12);

\path[draw=drawColor,line width= 0.4pt,line join=round,line cap=round] (119.09,137.14) circle (  1.12);

\path[draw=drawColor,line width= 0.4pt,line join=round,line cap=round] (119.09,137.47) circle (  1.12);

\path[draw=drawColor,line width= 0.4pt,line join=round,line cap=round] (119.09,142.78) circle (  1.12);

\path[draw=drawColor,line width= 0.4pt,line join=round,line cap=round] (119.09,163.53) circle (  1.12);

\path[draw=drawColor,line width= 0.4pt,line join=round,line cap=round] (119.09,157.13) circle (  1.12);

\path[draw=drawColor,line width= 0.4pt,line join=round,line cap=round] (119.09,136.35) circle (  1.12);

\path[draw=drawColor,line width= 1.2pt,line join=round] (134.85, 92.20) -- (155.87, 92.20);

\path[draw=drawColor,line width= 0.4pt,dash pattern=on 4pt off 4pt ,line join=round,line cap=round] (145.36, 68.08) -- (145.36, 85.20);

\path[draw=drawColor,line width= 0.4pt,dash pattern=on 4pt off 4pt ,line join=round,line cap=round] (145.36,108.87) -- (145.36, 97.64);

\path[draw=drawColor,line width= 0.4pt,line join=round,line cap=round] (140.11, 68.08) -- (150.62, 68.08);

\path[draw=drawColor,line width= 0.4pt,line join=round,line cap=round] (140.11,108.87) -- (150.62,108.87);

\path[draw=drawColor,line width= 0.4pt,line join=round,line cap=round] (134.85, 85.20) --
	(155.87, 85.20) --
	(155.87, 97.64) --
	(134.85, 97.64) --
	(134.85, 85.20);

\path[draw=drawColor,line width= 0.4pt,line join=round,line cap=round] (145.36, 65.64) circle (  1.12);
\end{scope}
\begin{scope}
\path[clip] (  0.00,  0.00) rectangle (187.90,216.81);
\definecolor{drawColor}{RGB}{0,0,0}

\path[draw=drawColor,line width= 0.4pt,line join=round,line cap=round] ( 66.54, 61.20) -- (145.36, 61.20);

\path[draw=drawColor,line width= 0.4pt,line join=round,line cap=round] ( 66.54, 61.20) -- ( 66.54, 55.20);

\path[draw=drawColor,line width= 0.4pt,line join=round,line cap=round] ( 92.81, 61.20) -- ( 92.81, 55.20);

\path[draw=drawColor,line width= 0.4pt,line join=round,line cap=round] (119.09, 61.20) -- (119.09, 55.20);

\path[draw=drawColor,line width= 0.4pt,line join=round,line cap=round] (145.36, 61.20) -- (145.36, 55.20);

\path[draw=drawColor,line width= 0.4pt,line join=round,line cap=round] ( 49.20, 74.10) -- ( 49.20,156.26);

\path[draw=drawColor,line width= 0.4pt,line join=round,line cap=round] ( 49.20, 74.10) -- ( 43.20, 74.10);

\path[draw=drawColor,line width= 0.4pt,line join=round,line cap=round] ( 49.20, 94.64) -- ( 43.20, 94.64);

\path[draw=drawColor,line width= 0.4pt,line join=round,line cap=round] ( 49.20,115.18) -- ( 43.20,115.18);

\path[draw=drawColor,line width= 0.4pt,line join=round,line cap=round] ( 49.20,135.72) -- ( 43.20,135.72);

\path[draw=drawColor,line width= 0.4pt,line join=round,line cap=round] ( 49.20,156.26) -- ( 43.20,156.26);

\node[text=drawColor,rotate= 90.00,anchor=base,inner sep=0pt, outer sep=0pt, scale=  1.00] at ( 34.80, 74.10) {4};

\node[text=drawColor,rotate= 90.00,anchor=base,inner sep=0pt, outer sep=0pt, scale=  1.00] at ( 34.80, 94.64) {6};

\node[text=drawColor,rotate= 90.00,anchor=base,inner sep=0pt, outer sep=0pt, scale=  1.00] at ( 34.80,115.18) {8};

\node[text=drawColor,rotate= 90.00,anchor=base,inner sep=0pt, outer sep=0pt, scale=  1.00] at ( 34.80,135.72) {10};

\node[text=drawColor,rotate= 90.00,anchor=base,inner sep=0pt, outer sep=0pt, scale=  1.00] at ( 34.80,156.26) {12};
\end{scope}
\begin{scope}
\path[clip] (  0.00,  0.00) rectangle (187.90,216.81);
\definecolor{drawColor}{RGB}{0,0,0}

\node[text=drawColor,anchor=base,inner sep=0pt, outer sep=0pt, scale=  1.00] at (105.95, 15.60) {Scenario C};
\end{scope}
\begin{scope}
\path[clip] (  0.00,  0.00) rectangle (187.90,216.81);
\definecolor{drawColor}{RGB}{0,0,0}

\path[draw=drawColor,line width= 0.4pt,line join=round,line cap=round] ( 49.20, 61.20) --
	(162.70, 61.20) --
	(162.70,167.61) --
	( 49.20,167.61) --
	( 49.20, 61.20);

\path[draw=drawColor,line width= 0.4pt,line join=round,line cap=round] ( 66.54, 61.20) -- (145.36, 61.20);

\path[draw=drawColor,line width= 0.4pt,line join=round,line cap=round] ( 66.54, 61.20) -- ( 66.54, 55.20);

\path[draw=drawColor,line width= 0.4pt,line join=round,line cap=round] ( 92.81, 61.20) -- ( 92.81, 55.20);

\path[draw=drawColor,line width= 0.4pt,line join=round,line cap=round] (119.09, 61.20) -- (119.09, 55.20);

\path[draw=drawColor,line width= 0.4pt,line join=round,line cap=round] (145.36, 61.20) -- (145.36, 55.20);

\node[text=drawColor,anchor=base,inner sep=0pt, outer sep=0pt, scale=  0.80] at ( 66.54, 39.60) {$\mathcal{I}_{v}$};

\node[text=drawColor,anchor=base,inner sep=0pt, outer sep=0pt, scale=  0.80] at ( 92.81, 39.60) {$\mathcal{I}_{c}$};

\node[text=drawColor,anchor=base,inner sep=0pt, outer sep=0pt, scale=  0.80] at (119.09, 39.60) {$\mathcal{I}_{e}$};

\node[text=drawColor,anchor=base,inner sep=0pt, outer sep=0pt, scale=  0.80] at (145.36, 39.60) {$\mathcal{I}_{p}$};
\end{scope}
\end{tikzpicture}
\end{center}
 \caption{Box plots of the logarithm of the widths of the confidence intervals.}
\label{fig:box}	
\end{figure}

We also calculate the average widths of the four interval estimators.
Figure~\ref{fig:box} displays the box plots of the logarithm of
these average widths for scenarios A--C with $N_0=200$ and $K=2$.
Compared with the EL interval $\mI_e$,
the PEL interval $\mI_p$ has much narrower average widths
although they have almost the same coverage probabilities.
The CL intervals $\mI_c$ and $\mI_e$
also have narrower average widths than
the EL interval $\mI_e$. However, they have much lower coverage probabilities than the latter.
In addition, the width of the PEL interval $\mI_p$
has the lowest dispersion of the four intervals.

In summary, the proposed EM algorithm produces more reliable
results than the standard optimization algorithm.
The PEL point estimator calculated with the proposed EM algorithm
has much more stable performance than the other estimators.
The corresponding PEL intervals are much narrower than
the EL intervals with nearly no loss of coverage probabilities
and with the most stable widths.

\section{Real-world data analysis}
\label{sec:data}

In this section, we analyze a real-world data set,
named  black bear data,
to demonstrate the advantages of the proposed
 PEL estimation method and the proposed EM algorithm.


To estimate the abundance of black bears
at the military installation Fort Drum in northern New York, USA,
data on black bears were collected over 8 weeks during June and July 2006
\citep{gardner2010estimating, royle2013spatial}.
Although the survey was conducted using 38 baited traps,
we integrate the capture histories of 47 individuals
and treat this data set as discrete-time
capture--recapture data. Besides the encounter histories,
the covariate sex is also available for the bears caught.
We analyze the data with the CL, EL, and PEL estimation methods
for the capture probability models $\rM_{hb}$ and $\rM_{htb}$.
Table~\ref{tab:bears} tabulates the estimates of the abundance of black bears.
The PEL and EL methods were implemented by the proposed EM algorithm ({\tt R} package {\tt Abun})
and the CL method was implemented by
the optimization algorithm ({\tt R} package {\tt VGAM})
or the proposed EM algorithm.

\begin{table}[bt!]
\centering
\caption{Estimates for black bear abundances.$^\dagger$}
\label{tab:bears}
\begin{threeparttable}
\begin{tabular}{cccccccc}
	\toprule
	&&
  \multicolumn{3}{c}{Model $\rM_{hb}$}&
  \multicolumn{3}{c}{Model $\rM_{htb}$}
\\
\cmidrule(lr){3-5} \cmidrule(lr){6-8}
Algorithm & Method  & Est.	 & SE  & CI  & Est.	 & SE  & CI \\ 	
\midrule
EM & PEL  & 65 & 14.52 & [50, 165] & 106 & 111.37 & [51, 295] \\
EM & EL   & 65 & 14.54 & [50, 226] & 257 & 947.17 & [52, --]$^\ddagger$ \\
EM & CL   & 70 & 18.75 & [34, 107] & 949 & $1 \times 10^5$ & $[-3 \times 10^4, 3 \times 10^4]$ \\
Optimization & CL  & 70 & 18.53 & [34, 107] & $7 \times 10^8$ & $7 \times 10^{10}$ & $[-1 \times 10^{12}, 1 \times 10^{12}]$ \\
\bottomrule
\end{tabular}

\begin{tablenotes}
\footnotesize
\item[$\dagger$]
Est.: point estimate,
SE: standard error, and
CI: confidence interval at the 95\% confidence level.
\item[$\ddagger$] --: A number greater than $10^9$.
\end{tablenotes}

\end{threeparttable}  	
\end{table}

For model $\rM_{hb}$, the PEL and EL methods produce the same point estimates
and lower bounds of interval estimates,
and nearly identical standard errors.
This is probably because the common point estimate 65 is close to \cite{chao1987estimating}'s lower bound 63, and the penalty is hardly applied.
Even so, the PEL interval has a much smaller upper limit
and, hence, a much narrower width than the EL interval.
With the CL method, the two algorithms produce
almost the same point estimates, standard errors, and Wald confidence intervals.

The results are totally different for the most general $\rM_{htb}$ model.
The most stable and reasonable results are those from the PEL method.
The PEL abundance estimate is about 106 with
a standard error of 111.37, whereas the PEL interval estimate is [51, 295].
The point estimate of 106 is close to the 114 of \cite{gardner2010estimating}, which was also based on spatial information.
In contrast, the EL method produces rather unstable results.
The point estimate (257) and standard error (947.17) are both much larger,
and the interval estimate has a nearly infinite upper limit.
This comparison implies that the penalty in the PEL method is applied.
The results for the CL method are too unstable to be acceptable,
whether they are calculated by the proposed EM algorithm
or the optimization algorithm.

\begin{figure}
\begin{center}
\includegraphics[width=0.5\textwidth]{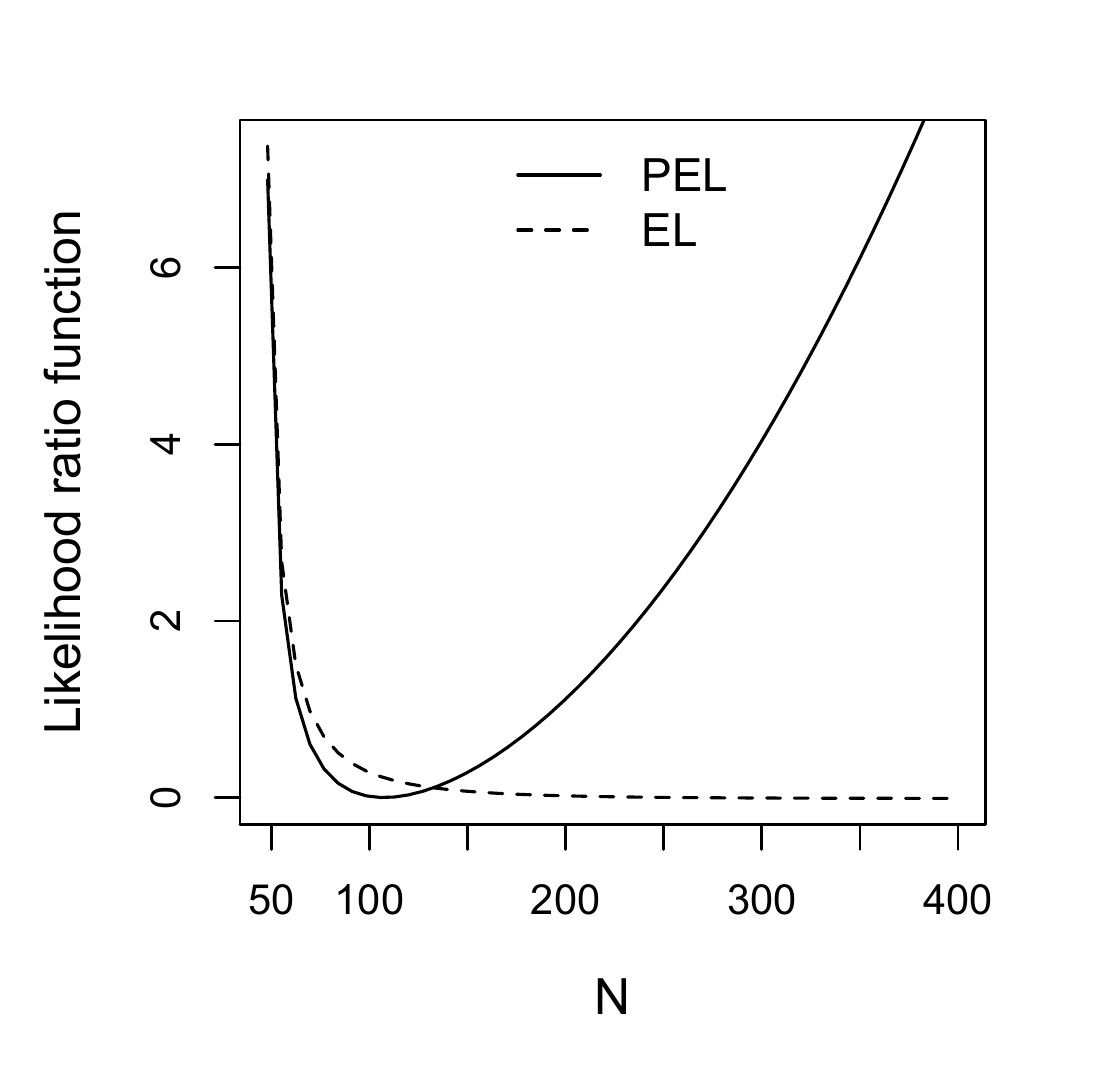}
\end{center}
 \caption{PEL (solid line) and EL (dashed line) ratio functions of $N$
 for the black bear data.}
\label{fig:plot-elr}
\end{figure}

To gain insights about the remarkable difference between
the PEL and EL methods for model $\rM_{htb}$,
we display the PEL and EL ratio functions of $N$ in
Figure~\ref{fig:plot-elr}.
It is clear that the EL ratio function
is decreasing and becomes flat for large $N$, which explains the undesirable
poor performance of the EL method.
With the recommended penalty, the PEL ratio function increases quickly for $N>150$.
Therefore, the PEL method successfully overcomes the instability of the EL method
and produces better and reliable point and interval estimates.
For model selection diagnostics, we apply the PEL-based Akaike information criterion (AIC) to
the goodness-of-fit of the probabilistic models.
The AICs of  models $\rM_{hb}$   and $\rM_{htb}$ are
829.33 and 828.73, respectively,
suggesting that   model $\rM_{htb}$ fits the data better than model $\rM_{hb}$.

\section{Conclusion and discussion}
\label{sec:con}

When the capture probability is moderate or low, 
the general capture--recapture model $\rM_{htb}$ may be weakly identified by the data
and the likelihood function of abundance may be so flat that
the estimation results may be  unstable;
see also Section \ref{sec:data}.
We compensate for the instability of model fitting
by penalizing large maximum EL estimates of abundance and
drawing them closer to  \cite{chao1987estimating}'s lower-bound estimate,
which is known to be  stable.
The penalty has a similar effect to imposing an informative prior in a Bayesian setting,
and the result is naturally a better fit with narrower confidence intervals
as it makes use of more information.
There is a close relation between our recommended penalty function and
that in the penalized likelihood $\ell_3$
of \cite{Wang2005penalized}.
Both penalties are data-adaptive and the target parameters have a quadratic form.
The difference is that \cite{Wang2005penalized}'s penalty is added to a CL
of an odd parameter, namely $\alpha/(1 - \alpha)$ in our notation,
whereas our penalty is added to an EL of the abundance.
We could use other penalty functions in the PEL method, such as
$f(N) = N$ or $-\log(N)$.
Our simulation experience shows that with these penalties,
the resulting PEL estimators are somewhat sensitive to
the choice of tuning parameters.
Also, the non-concavity of $-\log(N)$ would make calculating the PEL more challenging.

We propose to  implement
the PEL method by the EM algorithm,
which was proposed by  \cite{liu2022semiparametric}
under one-inflated capture--recapture models.
The EM algorithm guarantees that the likelihood
increases  after each iteration and
that the final estimator is equal to the maximum likelihood estimator.
Alternatively, the PEL method can be implemented
in a full Bayesian framework where the impact of the prior is transparent.
To investigate the behavioral effect of individuals on captures,
we consider an enduring (long-term) memory of the behavior
in the capture--recapture model $\rM_{htb}$
which means that
after an individual is captured,
the individual has a long memory of its first-capture experience and
the effect lasts in  the remaining period of the experiment.
In practice, ephemeral (short-term) behaviors are also frequently seen,
which means that the capture probability may depend on whether or not
it is caught on the most recent occasion
\citep{yang2005modeling, bartolucci2007class}.
The proposed PEL method and EM algorithm are both applicable to such cases,
as noted in Section~2 of the supplementary material.

There may   not be enough information in conventional capture--recapture data
to fit the demanding probability models reliably.
Recently, biologists have focused on using sampling designs that do deliver better information.
A prime example of this is the burgeoning field of spatial capture--recapture,
where individual heterogeneity is attributed to an animal's spatial location
relative to the traps, and the spatial information in the data is used in the fitting of the model.
It's notable that the real-data analysis of black bears was actually
taken from a spatial capture-recapture study, but we discarded
 the spatial information and used a sex covariate instead.
It is of interest to extend the proposed methods to the complicated
spatial capture--recapture data.

\bigskip
\begin{center}
{\large\bf SUPPLEMENTARY MATERIAL}
\end{center}

\begin{description}

\item[Title:]
The supplementary material for
``Penalized empirical likelihood estimation and EM algorithms
for closed-population capture--recapture models"
contains proofs of all the theorems and propositions
and extends the PEL method 
to more general capture--recapture models
with ephemeral behavioral effect. (Abun$\_$supp.pdf)

\item[R-package for  Abun routine:]
This package contains the code to perform  the EL and PEL methods
by the proposed EM algorithms.
The package also contains the real-world data set analyzed in the article.
(Abun$\_$0.1-1.tar.gz)

\item[Black bear data set:]
Data set used in the illustration of the PEL method
and the EM algorithm  in Section~ \ref{sec:data}. (blackbear.txt)

\end{description}

\section*{Acknowledgements}
This research is supported by the China Postdoctoral Science Foundation (Grant 2020M681220),
the National Natural Science Foundation of China (12101239 and 12171157),
the State Key Program of National Natural Science Foundation of China (71931004 and 32030063),
the Natural Sciences and Engineering Research Council of Canada (RGPIN-2020-04964),
and the 111 project (B14019).

\bibliographystyle{Chicago} 

\bibliography{abun-ref}
\end{document}